\documentclass[ reprint,nofootinbib,amsmath,amssymb,onecolumn,aps]{revtex4-2}
\usepackage{appendix}
\usepackage{graphicx}
\usepackage{dcolumn}
\usepackage[colorlinks,linkcolor=magenta,anchorcolor=cyan,citecolor=blue]{hyperref}
\usepackage{physics}
\usepackage{bm}

\def\be{\begin{equation}}
\def\ee{\end{equation}}
\def\ba{\begin{eqnarray}}
\def\ea{\end{eqnarray}}

\def\dbar{{\mathchar '26\mkern -10mu\delta}}

              \def\.{\cdot}

\begin{document}
\title{Quadratic curvature corrections to 5-dimensional Kerr-AdS black hole thermodynamics by background subtraction method}
\author{Gerui Chen$^{1,2}$}
\email{t20152277@csuft.edu.cn}
\author{Xiyao Guo$^{2,3}$}
\email{xiyaoguo@mail.bnu.edu.cn}
\author{Xin Lan$^{2,3}$}
\email{xinlan@mail.bnu.edu.cn}
\author{Hongbao Zhang$^{2,3}$}
\email{hongbaozhang@bnu.edu.cn}
\author{Wei Zhang$^{2,3}$}
\email{w.zhang@mail.bnu.edu.cn}
\affiliation{$^1$ College of Electronic Information and Physics, Central South University of Forestry and Technology, Changsha 410004, China\\
$^2$School of Physics and Astronomy, Beijing Normal University, Beijing 100875, China\\
$^3$ Key Laboratory of Multiscale Spin Physics, Ministry of Education, Beijing Normal University, Beijing 100875, China}

\date{\today}

\begin{abstract}
We justify the applicability of the background subtraction  method to both Einstein's gravity and its higher derivative corrections in  5-dimensional asymptotically AdS spacetimes, where the corresponding higher derivative corrections to the expression for the ADM mass and angular momentum are also worked out. Then we further apply the background subtraction method to calculate the first order corrected Gibbs free energy by the quadratic curvature terms for the 5-dimensional Kerr-AdS black hole, which is in exact agreement with the previous result obtained by the holographic renormalization method. Such an agreement in turn substantiates the applicability of the background subtraction method.

\end{abstract}
\maketitle
\section{Introduction}

In the Euclidean approach to black hole thermodynamics, there have been two methods proposed to obtain the finite Gibbs free energy of stationary black holes and the relevant thermodynamic quantities. One is the background subtraction method invented by Hawking and his companions with the inception of the Euclidean approach to quantum gravity \cite{Hartle,GH,HP,HH,HHT,GPP}, which has been used to successfully extract various higher derivative corrections to the static black hole thermodynamics \cite{GKT,GL,Landsteiner,CK,CNO,N1,N2,CN,DG}. 
The other is the covariant counterterm method, also dubbed as holographic renormalization method because it is motivated by Anti-de Sitter/Conformal Field Theory (AdS/CFT) correspondence \cite{BFS,PS,MM}. Both methods are believed to have their advantages and disadvantages, respectively. Although the holographic renormalization method is supposed to be universally applicable, the involved calculation is cubersome because it needs complicated dimension-dependent boundary terms in addition to the generalized Gibbons-Hawking-York term for a generic diffeomorphism covariant gravity theory. Compared to this, the background subtraction method is much simpler, but it is suspected to have a rather restrictive applicability in the sense that the induced boundary metric of the reference spacetime cannot be the exactly the same as that of stationary black holes. 
Gratefully, one very progress has been made most recently regarding the applicability of the background subtraction method in \cite{HB}, where it is found that the criterion for the applicability of the background subtraction method does not require the aforementioned induced boundary metrics be exactly the same and the background subtraction method turns out to be as applicable as the covariant counterterm method indeed. However, such an applicability is explicitly demonstrated only for the 4-dimensional rotating black hole. The purpose of this paper is two-fold. For one thing, we shall show that the background subtraction method is also applicable to the 5-dimensional Kerr-AdS black hole for the Einstein's gravity and its higher derivative corrections. For another, we shall apply the background subtraction method to calculate the quadratic curvature corrections to the 5-dimensional Kerr-AdS black hole thermodynamics. Among the recent efforts in the investigation of the higher derivative corrections to black hole thermodynamics \cite{RS,Cheung,Cremonini,Belgium,Melo,Bobev,Xiao1,Cassani,Noumi,Cassani1,Ma,Zatti,Hu,Ma2,Cano,Massai,Cassani2,Xiao2,HB1}, the authors in \cite{Ma2} have already obtained such quadratic curvature corrections to the Gibbs free energy for the 5-dimensional Kerr-AdS black hole. But the method they use is the holographic renormalization method. So this provides us with a very opportunity to compare our result with theirs. The resulting exact agreement between them further substantiates the applicability of the background subtraction method.

This paper will be structured as follows. After a brief review of the criterion for the applicability of the background subtraction method in the subsequent section, we shall demonstrate in an explicit way that such a criterion is satisfied by Einstein's gravity and its higher derivative corrections in 5-dimensional asymptotically AdS spacetimes in Section \ref{5dads}. Then in Section \ref{33}, we shall apply the background subtraction method to calculate the first order corrected Gibbs free energy by the quadratic curvature terms and compare it with that obtained previously by the holographic renormalization method.
We conclude our paper with some discussions in the final section. 

\section{Criterion for the applicability of background subtraction method in $F(R_{abcd})$ gravity }\label{bsm}
In this section, we present a brief review  of the criterion for the applicability of the background subtraction method in $F(R_{abcd})$ gravity \cite{HB}. As such, let us consider the Lagrangian form for $F(R_{abcd})$ gravity as follows
\begin{equation}
\mathbf{L}=\bm{\epsilon}F(R_{abcd},g_{ab}),
\end{equation}
where $\bm{\epsilon}$ is the spacetime volume element. The variation of the Lagrangian form can be cast into the following form
\begin{equation}\label{variationradius}
    \delta \mathbf{L}= \bm\epsilon E^{ab}\delta g_{ab}+d\mathbf{\Theta}.
\end{equation}
 Here
\begin{eqnarray}
E_g^{ab}=\frac{1}{2}g^{ab}F+\frac{1}{2}\frac{\partial F}{\partial g_{ab}}+2\nabla_{c}\nabla_{d}\psi^{c(ab)d}
\end{eqnarray}
with the equation of motion given by $E_g^{ab}=0$, 
and $\mathbf{\Theta}=\theta\cdot\bm{\epsilon}$ is called the bulk symplectic potential with 
 \begin{equation}
\theta^a=2(\nabla_d\psi^{bdca}\delta g_{bc}-\psi^{bdca}\nabla_d\delta g_{bc}),
\end{equation}
where $\psi^{abcd}$ is defined as the derivative of $F$ with respect to $R_{abcd}$, namely
$\psi^{abcd}\equiv \frac{\partial F}{\partial R_{abcd}}$.

Now let us fix a timelike boundary $\Gamma$ and denote its outward-pointing normal vector and the induced metric as $n_a$ and $h_{ab}$, respectively. Then we have $\delta n_a=\delta a n_a$. Accordingly, the variation of the metric on $\Gamma$ can be written as 
\begin{equation}
    \delta g^{ab}|_\Gamma=-2\delta an^an^b-\dbar{A}^an^b-\dbar{A}^bn^a+\delta h^{ab}
\end{equation}
with $n_a\dbar{A}^a=0$. Whence $\mathbf{\Theta}$ on $\Gamma$ can be reduced to the following expression
\begin{eqnarray}\label{boundaryex}
    \mathbf{\Theta}|_{\Gamma}=-\delta \mathbf{B}+d\mathbf{C}+\mathbf{F}.
\end{eqnarray}
Here
\begin{eqnarray}
    \mathbf{B}=4\Psi_{ab}K^{ab}\hat{\bm\epsilon},\quad
    \mathbf{C}=\mathbf\omega\cdot\hat{\bm\epsilon},\quad
    \mathbf{F}=\hat{\bm\epsilon}(T_{hbc}\delta h^{bc}+T_{\Psi bc}\delta\Psi^{bc})
\end{eqnarray}
with $K_{ab}$ the extrinsic curvature, $\hat{\bm\epsilon}$ the induced volume defined as  $\bm\epsilon=\mathbf n\wedge \hat{\bm\epsilon}$, $\Psi_{ab}=\psi_{acbd}n^cn^d$, and
\begin{eqnarray}\label{bc}
    \omega^a&=&-2\Psi^a{}_b\dbar A^b+2h^{ae}\psi_{ecdb}n^d\delta h^{bc},\nonumber\\
    T_{hbc}&=&-2\Psi_{de}K^{de}h_{bc}+2n^a\nabla^e\psi_{deaf}h^d{}_{(b}h^f{}_{c)}-2 \Psi_{a(b}K^a{}_{c)}-2D^a(h_a{} ^eh_{(c}{}^f\psi_{|efd|b)}n^d),\nonumber\\
    T_{\Psi bc}&=&4 K_{bc},
\end{eqnarray}
where $D_a$ represents the covariant derivative compatible with the induced metric on $\Gamma$.

As shown in \cite{HB} for the Killing vector field $\xi$ normal to the stationary black hole horizon and tangential to $\Gamma$, we have
   \begin{equation}
     T\delta S=\delta [H_\xi]+\int_\mathcal{S}\xi\cdot[\mathbf{F}]\label{somenew}
    \end{equation}
with $\mathcal{S}$ the intersection of $\Gamma$ and a hypersurface emanating from the bifurcation surface $\mathcal{B}$. Here the square bracket represents the difference from the reference spacetime, for instance, $[\mathbf{F}]\equiv \mathbf{F}-\mathbf{F}^0$. $T$ denotes the Hawking temperature for the black hole while $S$ represents the black hole entropy, given by 
    \begin{equation}\label{entropyformula}
        S=\beta\int_\mathcal{B}\mathbf{Q}_\xi
        \end{equation}
     with $\beta=\frac{1}{T}$ the period along the Euclidean time $\tau=it$,  and
    \begin{equation}
\mathbf{Q}_\xi=-\psi^{cadb}\nabla_{[d}\xi_{b]}\bm{\epsilon}_{ca\cdot\cdot\cdot}.
\end{equation}
      In addition, $H_\xi$ is the Hamiltonian conjugate to $\xi$ and can be expressed as
    \begin{equation}
    H_\xi=\int_\mathcal{S}q_\xi\cdot\hat{\bm{\epsilon}},
    \end{equation}
    where 
    \begin{equation}
        q_\xi^a=\mathcal{T}^a{}_c\xi^c
    \end{equation}
    with 
    $\mathcal{T}^a{}_c= -(2T_h{}^a{}_c+2\Psi^{ab}T_{\Psi cb})$ regarded as the generalized Brown-York boundary energy momentum tensor.

    Note that unlike the first term on the right handed side of Eq. (\ref{somenew}), the second term is not a complete variation, So to have our familiar first law of black hole thermodynamics, we require
    \begin{equation}\label{ourcriterion}
        \int_{\mathcal{S}}\xi\cdot[\mathbf{F}]=0  \quad \texttt{as}
   \quad \Gamma\rightarrow\infty.  \end{equation}
    
     This requirement together with the finiteness condition for $[H_\xi]$ as $\Gamma\rightarrow\infty$ constitutes the necessary and sufficient condition for the applicability of the background subtraction method. If such a criterion is satisfied, then the Gibbs free energy of the stationary black holes can be expressed as 
    \begin{equation}
        \beta G\equiv\beta([H_\xi]-TS)=[I_E],
    \end{equation}
    where the Euclidean action $I_E$ is given by
    \begin{equation}
        I_E=\int_M\mathbf{L}_E+\int_\infty \mathbf{B}_E
    \end{equation}
    with $\mathbf{L}_E=-i\mathbf{L}$ and $\mathbf{B}_E=-i\mathbf{B}$.

    \section{Applicability of background subtraction method to 5-dimensional asymptotically AdS spacetimes}\label{5dads}
 In this section, we shall show in an explicit way that the criterion developed in the previous section is satisfied by Einstein's gravity and its higher derivative corrections in 5-dimensional asymptotically AdS spacetimes. In passing, we also present the explicit correction to the ADM (Arnowitt-Deser-Misner) mass and angular momentum by the higher derivative terms. 
    
Let us first demonstrate that the above criterion is satisfied by Einstein's gravity in 5 dimension. The corresponding Lagrangian form is given by 
    \begin{equation}\label{EH}
        \mathbf{L}_{eh}=\frac{1}{16\pi}\bm{\epsilon}(R+\frac{12}{l^2})
    \end{equation}
    with $l$ the AdS radius.
    Whence one can obtain $\psi_{abcd}=\frac{1}{32\pi}(g_{ac}g_{bd}-g_{ad}g_{bc})$ and $\Psi_{ab}=\frac{1}{32\pi}h_{ab}$. Accordingly, we further have
    \begin{eqnarray}
        \mathbf{B}=\frac{K}{8\pi}\hat{\bm{\epsilon}},\quad \mathbf{C}=-\frac{1}{16\pi}\dbar{A}\cdot\hat{\bm{\epsilon}},
         \quad \mathbf{F}=-\frac{1}{2}T_{bc}\delta 
        h^{bc}\hat{\bm\epsilon},
       \quad q_\xi^a=T^a{}_c\xi^c,
    \end{eqnarray}
    where $T_{bc}= -\frac{1}{8\pi}(K_{bc}-Kh_{bc})$ is the well-known Brown-York boundary energy-momentum tensor. 
    On the other hand, the above Lagrangian form allows the rotating Kerr-AdS black hole solution, which was firstly obtained in \cite{HHT}. For our purpose, we like to write this solution as follows \cite{GPP}
 \begin{eqnarray}
        ds^2&=&-W(1+\frac{r^2}{l^2})dt^2+
        \frac{2m}{U}(Wdt-\frac{a\sin^2\theta d\phi}{\Xi_a}-\frac{b\cos^2\theta d\psi}{\Xi_b})^2+\frac{U}{V-2m}dr^2\nonumber\\
        &&+\frac{U}{W\Xi_a\Xi_b}d\theta^2+
        \frac{r^2+a^2}{\Xi_a}\sin^2\theta d\phi^2+
        \frac{r^2+b^2}{\Xi_b}\cos^2\theta d\psi^2
\end{eqnarray}
    in the Boyer-Lindquist coordinates, 
    where 
    \begin{eqnarray}
        &&W=\frac{\sin^2\theta}{\Xi_a}+\frac{\cos^2\theta}{\Xi_b}, \quad \Xi_a=1-\frac{a^2}{l^2},\quad \Xi_b=1-\frac{b^2}{l^2},\nonumber\\
        &&U=r^2+a^2\cos^2\theta+b^2\sin^2\theta,\quad V=\frac{1}{r^2}(1+\frac{r^2}{l^2})(r^2+a^2)(r^2+b^2)
    \end{eqnarray}
    with $0\le \theta\le\frac{\pi}{2}$ and $0\le\phi,\psi\le2\pi$.
    Then it is not hard to show that the Killing vector field normal to the black hole event horizon is given by 
    \begin{equation}
        \xi=\frac{\partial}{\partial t}+\frac{a(1+\frac{r_+^2}{l^2})}{r_+^2+a^2}\frac{\partial}{\partial\phi}+\frac{b(1+\frac{r_+^2}{l^2})}{r_+^2+b^2}\frac{\partial}{\partial\psi}, 
    \end{equation}
    whereby we can obtain the corresponding temperature and angular velocities for the black hole as follows
    \begin{equation}\label{T0}
    T=\frac{r_+}{2\pi}(1+\frac{r_+^2}{l^2})(\frac{1}{r_+^2+a^2}+\frac{1}{r_+^2+b^2})-\frac{1}{2\pi r_+},\quad \Omega_a=\frac{a(1+\frac{r_+^2}{l^2})}{r_+^2+a^2}, \quad 
    \Omega_b=\frac{b(1+\frac{r_+^2}{l^2})}{r_+^2+b^2},
\end{equation}
where $r_+$ denotes the radius of the black hole event horizon, satisfying $V(r_+)-2m=0$. 

 To proceed, we take the $r=\bar{r}$ surface as our timelike boundary $\Gamma$ and rescale our time coordinate as follows
\begin{equation}\label{shift1}
    t\rightarrow \frac{\sqrt{V(\bar{r})-2m}}{\sqrt{V(\bar{r})}}t
\end{equation}
for the solution with $m=0$, which corresponds virtually to the pure AdS solution in the Boyer-Lindquist coordinates and will be chosen as our reference spacetime. Accordingly,  the line element of our reference spacetime becomes
 \begin{eqnarray}
        ds^2&=&-W\frac{V(\bar{r})-2m}{V(\bar{r})}(1+\frac{r^2}{l^2})dt^2+\frac{U}{V}dr^2+\frac{U}{W\Xi_a\Xi_b}d\theta^2+
        \frac{r^2+a^2}{\Xi_a}\sin^2\theta d\phi^2+
        \frac{r^2+b^2}{\Xi_b}\cos^2\theta d\psi^2,
    \end{eqnarray}
whereby one can choose 
 \begin{eqnarray}
       e^0\equiv \sqrt{W\frac{V(\bar{r})-2m}{V(\bar{r})}(1+\frac{r^2}{l^2})}dt ,\quad &e^1&\equiv \sqrt{\frac{U}{V}}dr,\quad e^2\equiv \sqrt{\frac{U}{W\Xi_a\Xi_b}}d\theta,  \nonumber\\
       e^3\equiv\sqrt{\frac{r^2+a^2}{\Xi_a}}\sin\theta d\phi,&& \quad e^4\equiv\sqrt{\frac{r^2+b^2}{\Xi_b}}\cos\theta d\psi
    \end{eqnarray}
  as its orthonormal basis.  By a straightforward calculation, one can obtain the non-vanishing basis components of the induced metric and extrinsic curvature on $\Gamma$ as follows
\begin{eqnarray}
&& h_{00}=-1-\frac{2m(a^2b^2-a^2l^2\sin^2\theta-b^2l^2\cos^2\theta)}{l^2\Xi_a\Xi_b\bar{r}^4
}+\mathcal{O}(\frac{1}{\bar{r}^6}), \quad h_{03}=-\frac{2mal\sin\theta\sqrt{W}}{\sqrt{\Xi_a}\bar{r}^4}+\mathcal{O}(\frac{1}{\bar{r}^5}),\nonumber \\
&&h_{04}=-\frac{2mbl\cos\theta\sqrt{W}}{\sqrt{\Xi_b}\bar{r}^4}+\mathcal{O}(\frac{1}{\bar{r}^5}), \quad
h_{22}=1, \quad h_{33}=1+\frac{2ma^2\sin^2\theta}{\Xi_a\bar{r}^4}+\mathcal{O}(\frac{1}{\bar{r}^6}), \nonumber\\
&&h_{44}=1+\frac{2mb^2\cos^2\theta}{\Xi_b\bar{r}^4}+\mathcal{O}(\frac{1}{\bar{r}^6}),\quad h_{34}=\frac{2mab\sin\theta\cos\theta}{\sqrt{\Xi_a\Xi_b}\bar r^4}+\mathcal O(\frac{1}{\bar r^6}),\nonumber\\
&&K_{00}=-\frac{1}{l}+\frac{1}{2l\bar{r}^2}(l^2-{a^2\sin^2\theta}-b^2\cos^2\theta)+\frac{1}{8l^5\Xi_a\Xi_b\bar{r}^4}\{-l^2\Xi_a[b^6+b^4l^2-b^2l^2(5l^2+8m)+3l^4(l^2+8m)]\nonumber\\
&&+2(a^2-b^2)[-2a^4l^2\Xi_b+a^2(l^4-b^4)+l^2(b^4-2b^2l^2+l^4-8ml^2)]\sin^2\theta-3l^4(a^2-b^2)^2\Xi_a\Xi_b\sin^4\theta\}+\mathcal{O}(\frac{1}{\bar{r}^5}),\nonumber\\
&&
K_{03}=\frac{2ma\sin\theta\sqrt{W}}{\sqrt{\Xi_a} \bar{r}^4}+\mathcal{O}(\frac{1}{\bar{r}^5}),\quad K_{04}=\frac{2mb\cos\theta\sqrt{W}}{\sqrt{\Xi_b} \bar{r}^4}+\mathcal{O}(\frac{1}{\bar{r}^5}), \quad K_{34}=-\frac{2mab\sin\theta\cos\theta}{l\sqrt{\Xi_a\Xi_b}\bar r^4}+\mathcal O((\frac{1}{\bar r^5}),\nonumber\\
&& K_{22}=\frac{1}{l}+\frac{-2a^2+b^2+l^2+3(a^2-b^2)\sin^2\theta}{2l\bar{r}^2}+\frac{1}{8l\bar{r}^4}\{-l^2(8m+l^2)+2b^2l^2(1-3\sin^2\theta)+a^4(8-24\sin^2\theta+15\sin^4\theta),\nonumber\\
&&+b^4(-1-6\sin^2\theta+16\sin^4\theta)+a^2[2l^2(3\sin^2\theta-2)-b^2(4-30\sin^2\theta+30\sin^4\theta)]\}+\mathcal{O}(\frac{1}{\bar{r}^5}), \nonumber\\
&& K_{33}=\frac{1}{l}+\frac{l^2-a^2(1+\cos^2\theta)+b^2\cos^2\theta}{2l\bar{r}^2}-\frac{1}{8l^3\Xi_a\bar r^4}\{-l^2\Xi_a[8a^4-b^4+2b^2l^2-l^4-4a^2(b^2+l^2)-8ml^2]\nonumber\\
&&-2[4a^6-5a^4(b^2+l^2)-b^2l^2(b^2+l^2)+a^2(b^4+6b^2l^2+l^4-8ml^2)]\sin^2\theta-3l^2(a^2-b^2)^2\Xi_a\sin^4\theta\}+\mathcal{O}(\frac{1}{\bar{r}^5}),\nonumber\\
&& K_{44}=\frac{1}{l}+\frac{l^2+a^2\sin^2\theta-b^2(1+\sin^2\theta)}{2l\bar r^2}+\frac{1}{8l^3\Xi_b \bar r^4}\{3b^6-5b^4l^2+l^6+8ml^4+b^2l^2(l^2+8m)\nonumber\\
&&-2[-2a^4l^2\Xi_b +a^2(l^4-b^4)-b^2(b^4-2b^2l^2+l^4-8ml^2)]\sin^2\theta-3l^2(a^2-b^2)^2\Xi_b\sin^4\theta\}+\mathcal{O}(\frac{1}{\bar{r}^5}),
  \end{eqnarray}
whereby we can further obtain
\begin{eqnarray}\label{K5}
    K&=&\frac{4}{l}+\frac{-2a^2+b^2+l^2+3(a^2-b^2)\sin^2\theta}{2l\bar{r}^2}+\frac{[-2a^2+b^2+l^2+3(a^2-b^2)\sin^2\theta](a^2\cos^2\theta+b^2\sin^2\theta)}{l\bar{r}^4}\nonumber\\
    &&-\frac{1}{8l\bar{r}^6}\{-16a^6+8a^4(b^2+l^2)+2a^2l^4\Xi_b^2-(b^2+l^2)(b^4-2b^2l^2+l^4+8ml^2)+\nonumber\\
    &&(a^2-b^2)[56a^4-b^4+2bn^2l^2-l^4-16a^2(b^2+l^2)-8ml^2]\sin^2\theta-\nonumber\\
    &&3(a^2-b^2)^2[22a^2-3(b^2+l^2)]\sin^4\theta+25(a^2-b^2)^3\sin^6\theta\}+\mathcal{O}(\frac{1}{\bar{r}^7}),
  \end{eqnarray} 
as well as the non-vanishing basis components of the  Brown-York tensor of the 5-dimensional Kerr-AdS metric 
\begin{eqnarray}
&&T_{00}=-\frac{3}{8\pi l}+\frac{-3l^2+a^2(4-5\sin^2\theta)+b^2(-1+5\sin^2\theta)}{16\pi l\bar{r}^2}+\frac{1}{512\pi l^5\Xi_a\Xi_b\bar{r}^4}\{-47a^6b^2+22a^4b^2-47a^2b^6+49a^4b^2l^2\nonumber\\
&&+49a^2b^4l^2+47b^6l^2-71a^4l^4-2a^2b^2l^4-71b^4l^4+24l^8-448a^2b^2ml^2+128l^4m(a^2+b^2)+192ml^6-20(a^2-b^2)\nonumber\\
&&[-3b^4l^2+5b^2l^4-2l^6-3a^4l^2\Xi_b+a^2(3b^4-8b^2l^2+5l^4)+16ml^4]\cos2\theta-21l^4(a^2-b^2)^2\Xi_a\Xi_b\cos4\theta\}+\mathcal{O}(\frac{1}{\bar{r}^4}),\nonumber\\
&&T_{03}=-\frac{5ma\sin\theta\sqrt{W}}{4\pi\sqrt{\Xi_a} \bar{r}^4}+\mathcal{O}(\frac{1}{\bar{r}^5}),\quad T_{04}=-\frac{5mb\cos\theta\sqrt{W}}{4\pi\sqrt{\Xi_b} \bar{r}^4}+\mathcal{O}(\frac{1}{\bar{r}^5}), \quad T_{34}=\frac{5mab\sin\theta\cos\theta}{4\pi l\sqrt{\Xi_a\Xi_b}\bar r^4}+\mathcal O((\frac{1}{\bar r^5}),\nonumber\\
    && T_{22}=\frac{3}{8\pi l}+\frac{-2a^2+b^2+l^2+3(a^2-b^2)\sin^2\theta}{16\pi l\bar{r}^2}+\frac{1}{64\pi l \bar{r}^4}[8a^4+b^4-2b^2l^2+l^4-4a^2(b^2+l^2)+8ml^2-\nonumber\\
    &&2(a^2-b^2)(8a^2-b^2-l^2)\sin^2\theta +9(a^2-b^2)^2\sin^4\theta]+\mathcal{O}(\frac{1}{\bar{r}^5}),\nonumber\\
    &&T_{33}=\frac{3}{8\pi l}+\frac{-2a^2+b^2+l^2+5(a^2-b^2)\sin^2\theta}{16\pi l\bar{r}^2}-\frac{1}{64\pi l^3\Xi_a \bar{r}^4}\{-l^2\Xi_a[8a^4+b^4-2b^2l^2+l^4-4a^2(b^2+l^2)+8ml^2]+\nonumber\\
    &&[-32a^6+38a^4(b^2+l^2)+6b^2l^2(b^2+l^2)-2a^2(3b^4+22b^2l^2+3l^4+40ml^2)]\sin^2\theta-21l^2\Xi_a(a^2-b^2)^2\sin^4\theta\}+\mathcal{O}(\frac{1}{\bar{r}^5}),\nonumber\\
    &&T_{44}=\frac{3}{8\pi l}+\frac{-2b^2+a^2+l^2+5(b^2-a^2)\cos^2\theta}{16\pi l\bar{r}^2}-\frac{1}{64\pi l^3\Xi_a \bar{r}^4}[-16a^4b^4+8a^2b^4+3b^6+16a^4l^2-5b^4l^2-8a^2l^4+b^2l^4+\nonumber\\
    &&l^6+32mb^2l^2+8ml^4+40mb^2l^2\cos 2\theta+2l^2(a^2-b^2)\Xi_b(-18a^2+5b^2+3l^2)\sin^2\theta+21l^2(a^2-b^2)^2\Xi_b\sin^4\theta]+\mathcal{O}(\frac{1}{\bar{r}^5})\nonumber\\
\end{eqnarray}
with $m=0$ giving rise to the corresponding result for the reference  spacetime.  Although
$\sqrt{|h|}=\frac{\sqrt{(V-2m)U}r\sin\theta\cos\theta}{\Xi_a\Xi_b}$ 
is the same for the Kerr-AdS black hole and reference spacetime with the leading term given by $\frac{\bar{r}^4\sin\theta \cos \theta}{l\Xi_a\Xi_b}$, the induced metric of the Kerr-AdS black hole on $\Gamma$
is obviously different from that of the reference spacetime. But nevertheless, the background subtraction method turns out to be still applicable according to our criterion. 
To see this, we have the induced metrics
\begin{eqnarray}
 h'_{00}&=&-1-\frac{2(a\delta a \Xi_b^2\sin^2\theta+b\delta b\Xi_a^2\cos^2\theta)}{l^2\Xi_a^2\Xi_b^2 W}+\frac{1}{\bar{r}^4}\{\frac{2m(a^2b^2-a^2l^2\sin^2\theta-b^2l^2\cos^2\theta)}{l^2\Xi_a\Xi_b}+\frac{2}{l^{10}\Xi_a^3\Xi_b^3W}[l^{8}\Xi_a^2\Xi_b^2W(-2abm\nonumber\\
 &&(b\delta a+a\delta b)+\delta m l^4-\delta m l^2(a^2\cos^2\theta+b^2\sin^2\theta)+2ml^2(a\delta a\sin^2\theta+b\delta b\cos^2\theta))+l^2(ma^2b^2-a^2l^2\sin^2\theta-b^2l^2\cos^2\theta)\nonumber\\
 &&(4(-al^2\Xi_b\delta a+a^2b\delta b-bl^2\delta b)(l^2\Xi_a+b\delta b\sin^2\theta)+2l^4\Xi_a\Xi_b(a\delta a\cos^2\theta+b\delta b\sin^2\theta))]\}+\mathcal{O}(\frac{1}{\bar{r}^6}), \nonumber\\
 h'_{03}&=&-\frac{1}{\bar{r}^4}\{\frac{2mal\sin\theta\sqrt{W}}{\sqrt{\Xi_a}}+\frac{2l^3\sin\theta}{l^8\Xi_a^2\Xi_b^2\sqrt{W\Xi_a}}[(-ma^4l^2\Xi_b\delta a-4ma^3bl^2\delta b+4mabl^4\delta b+a^5(2mb\delta b-b^2\delta m+l^2\delta m ))\cos^2\theta\nonumber\\
 &&-l^2\Xi_b(2a^3l^2\delta m-l^4(m\delta a+2a\delta m)+(mb^2l^2\delta a-3ma^2l^2\Xi_b\delta a+2ab^2l^2\delta m-a^3b^2\delta m -a^3l^2\delta m)\sin^2\theta)]\}+\mathcal{O}(\frac{1}{\bar{r}^6}),\nonumber\\
 h'_{04}&=&-\frac{1}{\bar{r}^4}\{\frac{2mbl\cos\theta\sqrt{W}}{\sqrt{\Xi_b}}+\frac{2l^3\cos\theta}{l^8\Xi_a^2\Xi_b^2\sqrt{W\Xi_b}}[l^2(-2a^2+l^2)(-b^3\delta m+l^2(m\delta b+b\delta m))+(-6ma^2b^2l^2\delta b+3mb^2l^4\delta b+\nonumber\\
 &&a^4(3mb^2\delta b+ml^2\delta b-b^3\delta m+bl^2\delta m))\cos^2\theta+(2mabl^4\Xi_b^2\delta a+b^3l^2(-mb\delta b+b^2\delta m -l^2\delta m)+a^2(mb^4\delta b+ml^4\delta b\nonumber\\
 &&-b^5\delta m+bl^4\delta m))\sin^2\theta\}+\mathcal{O}(\frac{1}{\bar{r}^6}),\nonumber\\
 h'_{22}&=&1+\frac{2(a\delta a \cos^2\theta+b\delta b\sin^2\theta)}{l^2W\Xi_a\Xi_b}+\frac{2(a\delta a\cos^2\theta+b\delta b \sin^2\theta)}{\bar{r}^2}-\frac{2(a^2\cos^2\theta+b^2\sin^2\theta)(a\delta a\cos^2\theta+b\delta b\sin^2\theta)}{\bar r^4}+\mathcal{O}(\frac{1}{\bar{r}^6}),\nonumber\\
  h'_{33}&=&1+\frac{2a\delta a }{l^2\Xi_a}+\frac{2a\delta a}{ \bar{r}^2}+\frac{1}{\bar{r}^4}\{\frac{2ma^2\sin^2\theta}{\Xi_a}-\frac{2a}{l^4\Xi_a^2\Xi_b^2}[a^6\delta a-2a^4l^2\delta a-2ml^4\delta a \sin^2\theta+a^2l^2(l^2-2m\sin^2\theta)\delta a+a^3l^2\delta m\sin^2\theta\nonumber\\
  &&-al^4\delta m\sin^2\theta]\}+\mathcal{O}(\frac{1}{\bar{r}^6}),\nonumber\\
   h'_{44}&=&1+\frac{2b\delta b }{l^2\Xi_b}+\frac{2b\delta b}{ \bar{r}^2}+\frac{1}{\bar{r}^4}\{\frac{2mb^2\cos^2\theta}{\Xi_b}-\frac{2b}{l^4\Xi_a^2\Xi_b^2}[b^6\delta b-2b^4l^2\delta b-2ml^4\delta b \cos^2\theta+b^2l^2(l^2-2m\cos^2\theta)\delta b+b^3l^2\delta m\cos^2\theta\nonumber\\
  &&-bl^4\delta m\cos^2\theta]\}+\mathcal{O}(\frac{1}{\bar{r}^6}),\nonumber\\
     h'_{34}&=&\frac{1}{\bar{r}^4}\{\frac{2mab\sin\theta\cos\theta}{\sqrt{\Xi_a\Xi_b}}-\frac{2\sin\theta\cos\theta\sqrt{\Xi_a\Xi_b}}{l^6\Xi_a^2\Xi_b^2}[-a^2bl^2\Xi_b\delta m+mbl^4\Xi_b\delta a +a(a^2+l^2)(mb^2\delta b+ml^2\delta b-b^3\delta m+bl^2\delta m)]\}\nonumber\\
  &&+\mathcal{O}(\frac{1}{\bar{r}^6}),\nonumber\\
  h'^0_{00}&=&-1-\frac{ 2(a\Xi_b^2 \delta a\sin^2\theta+b\Xi_a^2\delta b\cos^2\theta)}{l^2\Xi_a^2\Xi_b^2W}+\frac{2\delta m l^2}{\bar{r}^4}+\mathcal{O}(\frac{1}{\bar{r}^6}), \quad
   h'^0_{03}=0, \quad
   h'^0_{04}=0,  \quad
   h'^0_{34}=0, \nonumber\\
   h'^0_{22}&=&1+\frac{2(a\delta a \cos^2\theta+b\delta b\sin^2\theta)}{l^2W\Xi_a\Xi_b}+\frac{2(a\delta a\cos^2\theta+b\delta b \sin^2\theta)}{\bar{r}^2}-\frac{2(a^2\cos^2\theta+b^2\sin^2\theta)(a\delta a\cos^2\theta+b\delta b\sin^2\theta)}{\bar r^4}+\mathcal O(\frac{1}{\bar r^6}),\nonumber\\
 h'^0_{33}&=&1+\frac{2a\delta a }{l^2\Xi_a }+\frac{2a\delta a}{ \bar{r}^2}+\frac{2a^3\delta a}{\bar r^4}+\mathcal{O}(\frac{1}{\bar{r}^6}), \quad 
 h'^0_{44}=1+\frac{2b\delta b }{l^2\Xi_b }+\frac{2b\delta b}{ \bar{r}^2}+\frac{2b^3\delta b}{\bar r^4}+\mathcal{O}(\frac{1}{\bar{r}^6}), 
\end{eqnarray}
for the first order perturbed Kerr-AdS black hole and  reference spacetime. By $\delta h^{ab}=-h^{ac}h^{bd}\delta h_{cd}$, one can further have
\begin{eqnarray}
 \delta{h'^{00}}&=&\frac{2(a\Xi_b^2\delta a\sin^2\theta+b\Xi_a^2\delta b\cos^2\theta)}{l^2\Xi_a^2\Xi_b^2W}-\frac{2}{l^4\Xi_a^2\Xi_b^2W\bar{r}^4}\{l^4\Xi_a^2(2mb\delta b+l^2\delta m)-2[-mal^4\Xi_b^2\delta a+bl^4(m\delta b+b\delta m)+a^4(mb\delta b\nonumber\\
 &&+l^2\delta m)-a^2l^2(2mb\delta b+b^2\delta m+l^2\delta m)]\sin^2\theta+(a^2-b^2)l^2\delta m\sin^4\theta\}+\mathcal{O}(\frac{1}{\bar{r}^5}), \nonumber\\
\delta{h'^{03}}&=&\frac{2l\sin\theta\sqrt W(m\delta a+a\delta m)}{\sqrt{\Xi_a}\bar{r}^4}+\mathcal{O}(\frac{1}{\bar{r}^5}),\quad
\delta{h'^{04}}=-\frac{2l\cos\theta\sqrt W(m\delta b+b\delta m)}{\sqrt{\Xi_b}\bar{r}^4}+\mathcal{O}(\frac{1}{\bar{r}^5}), \nonumber\\
\delta{h'^{34}}&=&\frac{2l^2\sin\theta\cos\theta}{r^4\sqrt{\Xi_a\Xi_b}}\{\frac{2ma^2b\delta a}{l^4\Xi_a}+\frac{2mab^2\delta b}{l^4\Xi_b}+\frac{1}{l^6\sqrt{\Xi_a\Xi_b}}[-m(a^2+l^2)bl^2\Xi_b\delta a]+a(a^2+l^2)(mb^2\delta b+ml^2\delta b-\nonumber\\
&&b^3\delta m+bl^2\delta m)\}+\mathcal{O}(\frac{1}{\bar{r}^5}), \nonumber\\
\delta{h'^{22}}&=&-\frac{2(a\delta a\cos^2\theta+b\delta b\sin^2\theta)}{l^2\Xi_a\Xi_bW}-\frac{2(a\delta a\cos^2\theta+b\delta b\sin^2\theta)}{\bar r^2}+\frac{2(a^2\cos^2\theta+b^2\sin^2\theta)(a\delta a\sin^2\theta+b\delta b\cos^2\theta)}{\bar r^4}+\mathcal{O}(\frac{1}{\bar{r}^5}), \nonumber\\
\delta{h'^{33}}&=&-\frac{2a\delta a}{l^2\Xi_a}-\frac{2a\delta a}{\bar{r}^2}-\frac{2a[-a^2\Xi_a\delta a+a(2m\delta a+a\delta m)\sin^2\theta]}{\Xi_a\bar{r}^4}+\mathcal{O}(\frac{1}{\bar{r}^5}),\nonumber\\
\delta{h'^{44}}&=&-\frac{2b\delta b}{l^2\Xi_b}-\frac{2b\delta b}{\bar{r}^2}-\frac{2b[-b^2\Xi_b\delta b+b(2m\delta b+b\delta m)\cos^2\theta]}{\Xi_b\bar{r}^4}+\mathcal{O}(\frac{1}{\bar{r}^5}),\nonumber\\
 \delta{h'^{0}}^{00}&=&\frac{2(a\Xi_b^2\delta a\sin^2\theta+b\Xi_a^2\delta b\cos^2\theta)}{l^2\Xi_a^2\Xi_b^2W}+\frac{2\delta m l^2}{\bar{r}^4}+\mathcal{O}(\frac{1}{\bar{r}^5}), \quad
\delta{h'^{0}}^{03}=0,\quad \delta{h'^{0}}^{04}=0, \nonumber\\
\delta{h'^{0}}^{22}&=&-\frac{2(a\delta a\cos^2\theta+b\delta b\sin^2\theta)}{l^2\Xi_a\Xi_bW}-\frac{2(a\delta a\cos^2\theta+b\delta b\sin^2\theta)}{\bar r^2}+\frac{2(a^2\cos^2\theta+b^2\sin^2\theta)(a\delta a\sin^2\theta+b\delta b\cos^2\theta)}{\bar r^4}+\mathcal{O}(\frac{1}{\bar{r}^5}), \nonumber\\
\delta{h'^{0}}^{33}&=&-\frac{2a\delta a}{l^2\Xi_a}-\frac{2a\delta a}{\bar{r}^2}-\frac{2a^3\delta a}{\bar r^4}+\mathcal{O}(\frac{1}{\bar{r}^5}), \quad
\delta{h'^{0}}^{44}=-\frac{2b\delta b}{l^2\Xi_b}-\frac{2b\delta b}{\bar{r}^2}-\frac{2b^3\delta b}{\bar r^4}+\mathcal{O}(\frac{1}{\bar{r}^5}).
\end{eqnarray}
Then a straightforward computation gives us that $\int_{\mathcal{S}}\xi\cdot[\mathbf{F}]$ is proportional to $\int_{0}^1dx(1-2x)=0$ with $x=\sin^2\theta$, which implies the validity of the first law of black hole thermodynamics. On the other hand, we have 
\begin{eqnarray}\label{pure}
&&T^t{}_t=\frac{3}{8\pi l}+\frac{-4a^2+b^2+3l^2+5(a^2-b^2)
\sin^2\theta}{16\pi l \bar{r}^2 }-\frac{1}{512\pi l^4\Xi_a\Xi_b \bar{r}^4}\{47a^6b^2-22a^4b^4+47a^2b^6-47a^6l^2-49a^4b^2l^2-\nonumber\\
&&49a^2b^4l^2-47b^6l^2+71a^4l^4+2a^2b^2l^4+71b^4l^4-24l^8+64ma^2b^2l^2+64ma^2l^4+64mb^2l^4-192ml^6+4(a^2-b^2)[-156b^4l^2\nonumber\\
&&+25b^2l^4-10l^6-15a^4l^2\Xi_b+5a^2(3b^4-8b^2l^2+5l^4)+32ml^4]\cos2\theta+21l^4\Xi_a\Xi_b(a^2-b^2)^2\cos4\theta\}+\mathcal{O}(\frac{1}{\bar{r}^5}),\nonumber\\
&&T^t{}_\phi=\frac{mal\sin^2\theta}{2\pi \Xi_a \bar{r}^4}+\mathcal{O}(\frac{1}{\bar{r}^5}), \quad {T^{t}}_\psi=\frac{mbl\cos^2\theta}{2\pi \Xi_b\bar r^4}+\mathcal O (\frac{1}{\bar{r}^5})
\end{eqnarray}
with $m=0$ giving rise to the corresponding result for the reference spacetime. Whence we obtain the familiar finite ADM mass and angular momenta as follows
\begin{equation}
    M\equiv [H_{\frac{\partial}{\partial t}}]=\frac{\pi m(2\Xi_a+2\Xi_b-\Xi_a\Xi_b)}{4\Xi_a^2\Xi_b^2},\quad J_a\equiv-[H_{\frac{\partial}{\partial\phi}}]=\frac{\pi ma}{2\Xi_a^2\Xi_b},\quad \quad J_b\equiv-[H_{\frac{\partial}{\partial\psi}}]=\frac{\pi mb}{2\Xi_a\Xi_b^2},
\end{equation}
which follows the finiteness of $[H_\xi]$. Thus the background subtraction method is applicable to Einstein's gravity in 5-dimensional asymptotically AdS spacetime.

Next let us examine the potential effect induced by the higher derivative terms composed of Ricci scalar $R$, the traceless part of Ricci tensor $\mathcal{R}_{ab}$ and Weyl tensor $C_{abcd}$. First, the AdS radius may get corrected by the higher derivative terms. To be more specific,  let us consider the Einstein-Hilbert Lagrangian form supplemented by the higher derivative terms like $\frac{\varepsilon}{16\pi}\bm{\epsilon}L_i$ with $i$ the total number of derivatives of metric and suppose that the pure AdS is also its solution. Then by evaluating Eq. (\ref{variationradius}) on top of the pure AdS with the perturbation induced by the scaling transformation $g_{ab}\rightarrow e^{2\lambda} g_{ab}$ for a small constant $\lambda$,  we end up with 
\begin{equation}\label{adsradius}
    3R+\frac{60}{l^2}+\varepsilon(5-i)L_i=0,
\end{equation}
where we have used $d\mathbf{\Theta}=0$ when evaluated on the pure AdS. Note that both $\mathcal{R}_{ab}$ and  $C_{abcd}$ vanish on the pure AdS, so the correction to the AdS radius comes solely from $R^n$ in the following way
\begin{equation}\label{adsradius}
    3R+\frac{60}{l^2}+\varepsilon(5-2n)R^n=0.
\end{equation}

To proceed, we assume that the leading form of the higher derivative corrected black hole solution is captured by the Kerr-AdS solution with the above corrected AdS radius. Accordingly, we find that the basis components of the extrinsic curvature and Riemann curvature of the Kerr-AdS solution display the following asymptotic behavior 
\begin{equation}
    K_{ab}=\mathcal{O}(1),\quad R=\mathcal{O}(1),\quad \mathcal{R}_{ab}=0,\quad C_{abcd}=\mathcal{O}(\frac{1}{\bar{r}^4}).
\end{equation}

For $L_{2n}=R^n$, we have
\begin{equation}\label{admc1}
    \mathbf{B}=\frac{\varepsilon}{8\pi}nR^{n-1}K\hat{\bm{\epsilon}},\quad \mathbf{F}=\varepsilon[-\frac{1}{2}nR^{n-1}T_{bc}\delta h^{bc}+\frac{1}{8\pi}n(n-1)KR^{n-2}\delta R]\hat{\bm{\epsilon}}=-\frac{\varepsilon}{2}nR^{n-1}T_{bc}\delta h^{bc}\hat{\bm{\epsilon}},
    \quad q_\xi^a=\varepsilon nR^{n-1}T^a{}_c\xi^c,
\end{equation}
where we have used the fact $R=-\frac{20}{l_e^2}$ with $l_e$ the corrected AdS radius through Eq. (\ref{adsradius}).
So for the higher derivative term $R^n$, the criterion for the applicability of the background subtraction method is obviously satisfied. But the corresponding ADM mass and angular momenta receive the corrections, which turn out to be  equally proportional to the expression for the pure Einstein's gravity. Furthermore, Eq. (\ref{K5}) implies no boundary contribution from $[\mathbf{B}_E]$ to the Gibbs free energy , which is the same as the case for the pure Einstein's gravity.

For $L_{2(n+1)}=R^nC^2$ with $C^2\equiv C_{abcd}C^{abcd}$, the potential contribution to $\mathbf{B}$ is given by
\begin{equation}
\mathbf{B}\sim\frac{\varepsilon}{4\pi}\Phi_{ab}K^{ab}\hat{\bm{\epsilon}}=\mathcal{O}(\frac{1}{\bar{r}^2})
\end{equation}
with $\Phi_{ab}\equiv 2R^nC_{acbd}n^cn^d$. So the corresponding boundary contribution to the Gibbs free energy also vanishes. 
In addition, the potential contribution to $\mathbf{F}$ is given by 
\begin{equation}
\mathbf{F}\sim\frac{\varepsilon}{8\pi}(-\Phi_{ab}K^a{}_c\delta h^{bc}+2K_{bc}\delta \Phi^{bc})\hat{\bm{\epsilon}}=\mathcal{O}(1)\propto dx(1-2x),
\end{equation}
thus the integral of $\mathbf{F}$ over $\mathcal{S}$ vanishes, implying the validity of the first law of black hole thermodynamics. On the other hand, the potential contribution to the generalized Brown-York boundary energy momentum tensor is given by 
\begin{equation}
    \mathcal{T}^a{}_c\sim-\frac{\varepsilon}{8\pi}(3\Phi^{ab}K_{cb}-K^{ab}\Phi_{cb})=\mathcal{O}(\frac{1}{\bar{r}^4}).
\end{equation}
To be more specific, we have 
\begin{eqnarray}
\label{admc2}
    \mathcal{T}^t{}_t&=&\frac{\varepsilon R^nm\{-l_e^2[3l_e^2+b^2(1-4\sin^2\theta)]+a^2[b^2+l_e^2(3-4\sin^2\theta)]\}}{\pi l_e^5\Xi_a\Xi_b \bar{r}^4} +\mathcal{O}(\frac{1}{\bar{r}^5}),\nonumber\\ \mathcal{T}^t{}_\phi&=&\frac{\varepsilon R^n4ma\sin^2\theta}{\pi l_e\Xi_a\bar{r}^4}+\mathcal{O}(\frac{1}{\bar{r}^5}),\quad \mathcal{T}^t{}_\psi=\frac{\varepsilon R^n4mb\cos^2\theta}{\pi l_e\Xi_b\bar{r}^4}+\mathcal{O}(\frac{1}{\bar{r}^5}),
\end{eqnarray}
whereby 
 the corresponding correction to the ADM mass and angular momenta can be obtained as follows
\begin{eqnarray}
    \delta M=\varepsilon\frac{ 8R^n M}{l_e^2}, \quad \delta J_a=\varepsilon\frac{ 8R^n J_a}{l_e^2}, \quad \delta J_b=\varepsilon\frac{ 8R^n J_b}{l_e^2},
\end{eqnarray}
all of which are also finite and equally proportional to the expression for the pure Einstein's gravity. Thus the higher derivative term $R^nC^2$ does not spoil the applicability of the background subtraction method. 

We conclude this section by pointing out that neither the applicability of the background subtraction method nor the expression for the ADM mass and angular momenta will be altered by any other foreseeable higher derivative term based on the order of magnitude estimate.

\section{Higher derivative corrections to 5-dimensional Kerr-AdS black hole thermodynamics}\label{33}
With the background subtraction method well justified in the previous section, 
we are now in a position to apply it to calculate the first order quadratic curvature corrections to 5-dimensional Kerr-AdS black hole thermodynamics. To be more precise, we shall consider the Einstein-Hilbert term (\ref{EH}) corrected by the following quadratic curvature terms\footnote{Here we have ignored the quadratic curvature term involving $\mathcal{R}_{ab}\mathcal{R}^{ab}$ due to the fact that $\mathcal{R}_{ab}=0$ for the Kerr-AdS metric. } 
\begin{equation}
  \mathbf{L}_{hd} = \frac{1}{16\pi}\bm{\epsilon}(\varepsilon_1L^2R^2+\varepsilon_2L^2C^2),
\end{equation}
where $L$ as some UV length scale has been introduced to ensure $\varepsilon_i$ dimensionless. 
First, according to Eq. (\ref{adsradius}), the AdS radius gets corrected in the following way
\begin{equation}\label{L}
l_e=l-\frac{10\varepsilon L^2}{3l}.
\end{equation}
Then by following the trick devised in \cite{Xiao2}, we like to decompose the full bulk action into the following two parts
\begin{equation}
  I'= \int_M( \mathbf{L}'_{eh}+\mathbf{L}'_{hd})
\end{equation}
with 
\begin{equation}
    \mathbf{L}'_{eh}=\frac{1}{16\pi}\bm{\epsilon}(R+\frac{12}{l_e^2}), \quad \mathbf{L}'_{hd}=\frac{1}{16\pi}\bm{\epsilon}(\frac{12}{l^2}-\frac{12}{l_e^2})+\mathbf{L}_{hd}.
\end{equation}
Accordingly, we can obtain the first order corrected Gibbs free energy by evaluating the corresponding bulk $[I'_E]$ on the Kerr-AdS metric and pure AdS metric with $l_e$ as the AdS radius. The underlying reason can be articulated as follows. 
First, as detailed in the previous section, there is no boundary contribution to the Gibbs free energy from $\mathbf{B}_E$. Second, Eq. (\ref{variationradius}) together with Eq. (\ref{boundaryex}) tells us that the contribution to the Gibbs free energy from $\mathbf{L}'_{eh}$ evaluated on the full first order corrected black hole solution is equal to that evaluated on the aforementioned Kerr-AdS metric, because their difference comes only from the boundary term $\int_\infty [\mathbf{F}_E]=\int_\infty d\tau\wedge\frac{\partial}{\partial\tau}\cdot \mathbf{F}_E=\beta\int_{\mathcal{S}}\xi\cdot[\mathbf{F}]=0$ with $\mathbf{F}_E=-i\mathbf{F}$\footnote{Here we are working with the grand canonical ensemble, so the full first order corrected black hole solution and the involved Kerr-AdS metric only with the AdS radius corrected share the same temperature and angular velocities.}. Third, the contribution to the first order corrected Gibbs free energy from $\mathbf{L}'_{hd}$ evaluated at the full first order corrected solution is also equal to that evaluated on the aforementioned Kerr-AdS metric, because their difference starts apparently from the second order of $\varepsilon_i$ with one $\varepsilon_i$ from the coefficients in $\mathbf{L}'_{hd}$ and the other $\varepsilon_i$ from the difference of the two metrics. 

Let us first evaluate the bulk contribution to $[I'_E]$ from $\mathbf{L}'_{eh}$. To this end, we have 
\begin{equation}
    \int_{BH} \mathbf{L}'_{ehE}  =\frac{\beta}{16\pi} \int_{0}^{2\pi} d\phi\int_{0}^{2\pi} d\psi\int_{0}^{\frac{\pi}{2}} d\theta \int_{r_+}^{\bar{r}}dr \frac{8}{l_e^2}\sqrt{|g|}=\frac{\beta \pi(\bar{r}^2-r_+^2)(\bar{r}^2+r_+^2+a^2+b^2)}{4l_e^2\Xi_a\Xi_b}
\end{equation}
and  
\begin{equation}\label{first}
   \int_{AdS} \mathbf{L}'_{ehE}=\frac{\beta}{16\pi} \int_{0}^{2\pi} d\phi\int_{0}^{2\pi} d\psi\int_{0}^{\frac{\pi}{2}} d\theta \int_{0}^{\bar{r}}dr \frac{8}{l_e^2}\sqrt{|g|}
   =\frac{\beta\pi}{4l_e^2\Xi_a\Xi_b}[\bar r^4+(a^2+b^2) \bar r^2-m l_e^2]+\mathcal O(\frac{1}{\bar{r}}), 
\end{equation}
where we have used $\sqrt{|g|}_{BH}=\frac{rU\sin\theta\cos\theta}{\Xi_a\Xi_b}$ and $\sqrt{|g|}_{AdS}=\frac{\sqrt{V(\bar{r})-2m}}{\sqrt{V(\bar{r})}}\frac{rU\sin\theta\cos\theta}{\Xi_a\Xi_b}$.
However, as pointed out in \cite{GPP}, the present coordinates  cannot cover the whole pure AdS for the case of $a,b\neq 0$. As such, we make the following coordinate transformation
\begin{eqnarray}
    y^2\Xi_a \cos^2\hat \theta=(r^2+a^2)\cos^2\theta, \quad  y^2\Xi_b \sin^2\hat \theta=(r^2+b^2)\sin^2\theta,
\end{eqnarray}
whereby the resulting metric for the pure AdS reads
\begin{eqnarray}
    ds^2=-\frac{V(\bar r)-2m}{V(\bar r)}(1+y^2l_e^{-2})dt^2+\frac{dy^2}{1+y^2l_e^{-2}}+y^2(d\hat \theta^2+\cos^2\hat \theta d\phi^2+ \sin^2\hat \theta d\psi^2),
\end{eqnarray}
 and the hypersurface $r=0$ is determined by 
\begin{eqnarray}
    \bar{y}^2(\frac{\Xi_a \cos^2\hat\theta}{a^2}+\frac{\Xi_b \sin^2\hat\theta}{b^2})=1.
\end{eqnarray}
So besides Eq. (\ref{first}), another contribution to $[I_E']$ from the pure AdS is given by the integral from $y=0$ to the above hypersurface, i.e.,
\begin{equation}
    \int_{AdS} \mathbf{L}'_{ehE}=\frac{\beta}{16\pi} \int_{0}^{2\pi} d\phi\int_{0}^{2\pi} d\psi\int_{0}^{\frac{\pi}{2}} d\hat{\theta }\int_{0}^{\bar{y}}dy \frac{8}{l_e^2}\sqrt{|g|}
   =\frac{\beta \pi a^2b^2}{4l_e^2\Xi_a\Xi_b}+\mathcal{O}(\frac{1}{\bar r^4}).
\end{equation}
Then it follows from the background subtraction that the resulting contribution to $[I'_E]$ can be written as
\begin{equation}
\frac{\beta \pi[m-l_e^{-2}(a^2+r_+^2)(b^2+r_+^2)]}{4\Xi_a\Xi_b}
\end{equation}
by taking $\bar{r}\rightarrow\infty$.

Furthermore, with $ C^2=\frac{96m^2(4r^2-U)(4r^2-3U)}{U^6}$ and by a similar calculation, we wind up with the first order corrected Gibbs free energy
\begin{eqnarray}\label{Gibbs}
    G&\simeq&\frac{\pi }{4\Xi_a\Xi_b }[m-l_e^{-2}(r_+^2+a^2)(r_+^2+b^2)]-10\varepsilon_1\frac{L^2}{l_e^2}\frac{\pi }{\Xi_a\Xi_b }[m-l_e^{-2}(r_+^2+a^2)(r_+^2+b^2)]\nonumber\\
    &&-\varepsilon_2
L^2\frac{\pi m^2[a^4b^4-6a^2b^2r_+^2(a^2+b^2)+r_+^4(a^4-20a^2b^2+b^4)+2r_+^6(a^2+b^2)+9r_+^8]}{\Xi_a\Xi_b(a^2+r_+^2)^3(b^2+r_+^2)^3},
\end{eqnarray}
where $\simeq$ means that the equation holds to the first order of $\varepsilon_i$. By viewing $r_+$, $a$, $b$, and $m$ as the function of $T$, $\Omega_a$ and $\Omega_b$ through Eq. (\ref{T0}) with $l$ replaced by $l_e$, we have the above Gibbs free energy as a function of temperature and angular velocities as it should be the case. 


It is noteworthy that 
the first order corrected Gibbs free energy has also been obtained by using the holographic renormalization method supplemented with the field redefinition in \cite{Ma2}. To  compare the two results, we are required to work with the same temperature and angular velocities. But the temperature and angular velocities in \cite{Ma2} are defined as Eq. (\ref{T0}) without correcting $l$. Thus
by requiring $\delta \Omega_a=0$, $\delta \Omega_b=0$,  $\delta T
=0$  and with the aid of (\ref{L}), we obtain
\begin{eqnarray}
a&=&\tilde{a}+\varepsilon_1 L^2\frac{20 \tilde{a}\tilde{r}_+^2[\tilde{a}^2\tilde{b}^2l^2+2\tilde{b}^2l^2\tilde{r}_+^2+(\tilde{a}^2+\tilde{b}^2-3l^2)\tilde{r}_+^4-2\tilde{r}_+^6]}{3l^2(\tilde{r}_+^2+l^2)[2\tilde{r}_+^6-\tilde{r}_+^4(\tilde{a}^2+\tilde{b}^2+l^2)+\tilde{a}^2\tilde{b}^2 l^2]},\nonumber\\
b&=&\tilde{b}+\varepsilon_1 L^2\frac{20 \tilde{b}\tilde{r}_+^2[\tilde{a}^2\tilde{b}^2l^2+2\tilde{a}^2l^2\tilde{r}_+^2+(\tilde{a}^2+\tilde{b}^2-3l^2)\tilde{r}_+^4-2\tilde{r}_+^6]}{3l^2(\tilde{r}_+^2+l^2)[2\tilde{r}_+^6-\tilde{r}_+^4(\tilde{a}^2+\tilde{b}^2+l^2)+\tilde{a}^2\tilde{b}^2 l^2]},\nonumber\\
r_+&=&\tilde{r}_++\varepsilon_1 L^2\frac{20  \tilde{r}_+^5(\tilde{a}^2+\tilde{b}^2-2\tilde{r}_+^2)}{3l^2[2\tilde{r}_+^6-\tilde{r}_+^4(\tilde{a}^2+\tilde{b}^2+l^2)+\tilde{a}^2\tilde{b}^2 l^2]},
\end{eqnarray}
where $\tilde{r}_+$, $\tilde{a}$ and $\tilde{b}$ denote the corresponding quantities in the Kerr-AdS metric with the AdS radius uncorrected. 
By substituting the above equation together with Eq. (\ref{L}) as well as $2m=V(r_+)$ into Eq.  (\ref{Gibbs}), we find that the resulting expression is in exact agreement with that obtained in \cite{Ma2} as it should be the case.

\section{Conclusion}
After justifying the applicability of the background subtraction method to 5-dimensional asymptotically AdS spacetimes, we apply it to calculate out the first order quadratic curvature corrected Gibbs free energy, in exact agreement with that previously obtained by the holographic renormalization method. It is noteworthy that we are not required to solve the full first order corrected black hole solutions by the quadratic curvature terms when calculating the first order corrected Gibbs free energy. But as demonstrated in \cite{HB1}, if interested one can also extract some generic information about the first order corrected black hole solutions by calculating the mass, angular momentum, and entropy from the Gibbs free energy through the standard thermodynamic relation and comparing the results with those obtained from the ADM and Wald formulas in the Lorentz signature. We have not presented such a result here because it turns out to be extraordinarily lengthy. 

As pointed out in the introduction of this paper, the background subtraction method is often blamed to be rather restrictive compared to the holographic renormalization method, albeit much simpler than the latter. The result obtained in  the present paper further substantiates the claim made in \cite{HB} that the background subtraction method is actually as applicable as the holographic renormalization method. 


Among others, there are two interesting issues worthy of further investigation. One is the first order corrections to the Gibbs free energy for our Kerr-AdS black hole by the higher derivative terms than the quadratic curvature ones. Note that with the background subtraction method, the whole calculation is supposed to be straightforward without any obstacle. This is different from the corresponding calculation by the holographic renormalization method, where one is required to introduce the additional counterterm for each higher derivative term. The other is to justify the applicability of the background subtraction method to more generic scenarios. So far, such an applicability has been demonstrated only for pure gravity. Although it is believed that such an applicability should hold in more generic situations, such as gravity coupled with matters, it is important to demonstrate this in an explicit manner as done in the present paper and \cite{HB,HB1} as well.  We hope to report our investigation along this line elsewhere in the future. 

\begin{acknowledgments}
This work is partially supported by the National Key Research and Development Program of China with Grant No. 2021YFC2203001 as well as the National Natural Science Foundation of China with Grant No. 12361141825, 12475049, and 12575047.

\end{acknowledgments}

\end{document}